\documentclass[iop]{emulateapj} 

\shorttitle{Companion to $\mu$~Her}
\shortauthors{Roberts et al.}         
 
\begin{document}


\title{CHARACTERIZATION OF THE COMPANION TO $\mu$~Her}

\author{Lewis C. Roberts, Jr.\altaffilmark{1},  
Brian D. Mason\altaffilmark{2},
Jonathan Aguilar\altaffilmark{3},
Joseph Carson\altaffilmark{4},
Justin Crepp\altaffilmark{5}, 
Charles Beichman\altaffilmark{1,6,7}, 
Douglas Brenner\altaffilmark{8}, 
Rick Burruss\altaffilmark{1},
Eric Cady\altaffilmark{1}, 
Statia Luszcz-Cook\altaffilmark{8}, 
Richard Dekany\altaffilmark{6}, 
Lynne Hillenbrand\altaffilmark{6},
Sasha Hinkley\altaffilmark{9},
David King\altaffilmark{10}, 
Thomas G. Lockhart\altaffilmark{1}, 
Ricky Nilsson\altaffilmark{8,11}, 
Rebecca Oppenheimer\altaffilmark{8}, 
Ian R. Parry\altaffilmark{10}, 
Laurent Pueyo\altaffilmark{3,12}, 
Emily L. Rice\altaffilmark{13}, 
Anand Sivaramakrishnan\altaffilmark{12}, 
R\'emi Soummer\altaffilmark{12},
Gautam Vasisht\altaffilmark{1}, 
Aaron Veicht\altaffilmark{8}, 
Ji Wang\altaffilmark{14},
Chengxing Zhai\altaffilmark{1},
Neil T. Zimmerman\altaffilmark{15}
} 

\altaffiltext{1}{Jet Propulsion Laboratory, California Institute of Technology, 4800 Oak Grove Drive, Pasadena CA 91109, USA}
\altaffiltext{2}{U.S. Naval Observatory, 3450 Massachusetts Avenue, NW, Washington, DC 20392-5420, USA}
\altaffiltext{3}{Johns Hopkins University, 3400 N. Charles Street, Baltimore, MD 21218, USA}

\altaffiltext{4}{Department of Physics \& Astronomy, College of Charleston, 58 Coming Street, Charleston, SC 29424, USA}
\altaffiltext{5}{Department of Physics, University of Notre Dame, 225 Nieuwland Science Hall, Notre Dame, IN, 46556, USA}\
\altaffiltext{6}{Division of Physics, Mathematics, and Astronomy, California Institute of Technology, Pasadena, CA 91125, USA}
\altaffiltext{7}{NASA Exoplanet Science Institute, 770 S. Wilson Avenue, Pasadena, CA 911225, USA}
\altaffiltext{8}{American Museum of Natural History, Central Park West at 79th Street, New York, NY 10024, USA}
\altaffiltext{9}{School of Physics, University of Exeter, Stocker Road, Exeter, EX4 4QL, UK}
\altaffiltext{10}{Institute of Astronomy, University of Cambridge, Madingley Rd., Cambridge, CB3 OHA, UK}
\altaffiltext{11}{Department of Astronomy, Stockholm University, AlbaNova University Center, Roslagstullsbacken 21, SE-10691 Stockholm, Sweden}
\altaffiltext{12}{Space Telescope Science Institute, 3700 San Martin Drive, Baltimore, MD 21218}

\altaffiltext{13}{Department of Engineering Science and Physics, College of Staten Island, City University of New York, Staten Island, NY 10314, USA}
\altaffiltext{14}{Department of Astronomy, Yale University, New Haven, CT 06511, USA}
\altaffiltext{15}{Princeton University, MAE, D207 Engineering Quad, Princeton, NJ 08544}



\begin{abstract}

$\mu$~Her is a nearby quadruple system with a G-subgiant primary and several low mass companions arranged in a 2$+2$ architecture. While the BC components have been well characterized, the Ab component has been detected astrometrically and with direct imaging but there has been some confusion over its nature, in particular whether the companion is stellar or substellar.  Using near-infrared spectroscopy we are able to estimate the spectral type of the companion as a M4$\pm$1V star.  In addition, we have measured the astrometry of the system for over a decade.  We combined the astrometry with archival radial velocity measurements to compute an orbit of the system.   From the combined orbit, we are able to compute the mass sum of the system.  Using the estimated mass of the primary, we  estimate the mass of the secondary as 0.32 M$_\sun$, which agrees with the estimated spectral type. Our computed orbit is preliminary due to the incomplete orbital phase coverage, but it should be sufficient to predict ephemerides over the next decade.   

\end{abstract}

  


\section{INTRODUCTION}

\object[HD 161797]{$\mu$~Her} (HD 161797 = HIP 86974 = WDS 17465+2743) is the third closest quadruple star system to the Sun \citep{davison2014} at a distance of 8.3 pc \citep{vanLeeuwen2007}.  The system consists of the G5IV primary \citep{morgan1973} surrounded by several M dwarfs. The closest companion is the Ab component, which has long been known from astrometric variations of the primary \citep{heintz1987} and from radial velocity (RV) variations \citep{cochran1987}.    The companion was first imaged by \citet{turner2001} with the adaptive optics (AO) system on the 2.5 m telescope at Mt.\ Wilson Observatory. There is also the BC pair, which is separated from the primary by 35\arcsec~and is a pair of M dwarfs in a 43.127$\pm$0.013 yr period \citep{prieur2014}. Both \citet{gould2004} and \citet{raghavan2010} concluded that the pair was physically bound to $\mu$~Her due to common proper motion.   There is also a wider D component with a separation of several hundred arcseconds, which is also an M dwarf.  \citet{raghavan2010} concluded that this is not a physical companion because the proper motion differs dramatically from that of the primary.  

While binary stars are common, 50\%$\pm$4\% of F6-G2 stars have companions \citep{raghavan2010}, quadruples are much less common.  Quadruples come in two major catagories, 3$+$1 systems, where a fourth star orbits a triple system in a wide orbit and,  2+2 systems, which consist of two binaries orbiting a common center of mass.  $\mu$~Her is the latter type, which is the more common type \citep{raghavan2010}. \citet{riddle2015} found that 10\% of nearby solar-type stellar binaries were actually 2+2 quadruples and this is likely linked to stellar system formation processes.  The $\mu$ Her system is one of the best-studied 2+2 quadrupole systems; therefore it serves as an archetype for understanding stellar system formation. Improving the characterization of this system is the motivation for this analysis. 

The primary star in the system exhibits solar-like oscillations \citep{bonanno2008} and these have been used to measure the mass of the primary \citep{yang2010} using asteroseimology (See Section \ref{orbit_analysis}).  A mass determination from an orbital analysis will serve as a useful crosscheck between these two techniques and can be used to improve the theoretical modeling of observed oscillation frequencies \citep{huber2014}. 
 
The nature of the $\mu$~Her Ab companion has been discussed for over 20 years.  Most indications are that it is an M-dwarf, but there are persistent possibilities that it is substellar.   \citet{torres1999} used a Monte Carlo technique to compute probability density functions of the mass of the close companion to $\mu$~Her~A based on limited RV data and hypothetical, future, single measurements of the angular separation. The study concluded that if the separation was 1\farcs4 then the companion would have a probable mass on the boundary between substellar and stellar. The angular separation measurements of \citet{turner2001} fell at the boundary between the two regimes.  From the computed $I$-band magnitude of the companion, they determined that the companion had to be redder than an M5V and could be substellar as \citet{torres1999} suggested.     \citet{debes2002} also observed the companion in the near-infrared with the Mt.\ Wilson AO system.  They were able to measure the photometry of the companion in $H$ and $K$ filters.  Using the  \textit{RI} photometry from \citet{turner2001} and their \textit{HK} photometry and model atmospheres they conclude that the mass should be $\approx$0.13 $M_\sun$ (equivalent to an M5V; \citealt{reid2005}).  They note that the \textit{RI} photometry is anomalously bright compared to the model, though the near-infrared photometry agreed with the models to within the error bars.  They concluded that the companion is stellar.  \citet{kenworthy2007} observed the system with the MMT in the mid-infrared and estimated the spectral class of the companion to be M4$\pm$1. Combining their results with that of \citet{debes2002}, they computed the (\textit{K}-\textit{M'}) color and noted that it was too red for an M4V star, but was similar to an early T-dwarf. With the possibility that the companion was a brown dwarf, we set out to resolve the prior conflicting color information and determine its spectral type with near-infrared spectroscopy.  We also collected multi-epoch astrometry to compute an resolved orbit for the system, allowing for the estimation of the individual masses of the objects.

\section{DATA}\label{observations}

\subsection{VISIM Observations}\label{visim_data}

We observed $\mu$~Her at the Advanced Electro-Optical System (AEOS) 3.6m telescope with the site AO system  and the Visible Imager (VisIm) camera \citep{roberts2002} on two separate dates: 2003 July 6 and  2005 April 24. Both data sets consist of 1000 frames  using a Bessel \textit{I}-band filter. After collection, any saturated frames were discarded and the remaining frames were debiased, dark subtracted and flat fielded.  The frames were weighted by their peak pixel, which is proportional to their Strehl ratio and then co-added using a shift-and-add routine. The resulting image was analyzed with the program \textit{fitstars} \citep{tenBrummelaar1996,tenBrummelaar2000}.  The 2005 image is shown in Figure \ref{images}b.

Error bars on the astrometry and photometry were assigned using the method in \citet{roberts2005}.   The extracted astrometry and differential photometry are given in Table \ref{astrometry}. The table lists the our new results as well as prior measures of the system.  The table gives the UT Besselian date of the observation, the measured position angle ($\theta$), the measured separation ($\rho$), the differential magnitude and the filter used for the observation. The final column is either the instrument that was used for the observation or a reference if it was a literature result. The table does not include null results from the literature. 

\subsection{Project 1640 Observations}

We used the Project 1640 (P1640) coronagraph \citep{hinkley2011} mounted on the PALM-3000 AO system \citep{dekany2013} at the Palomar 5 m telescope to observe $\mu$~Her three times in 2012 and 2015. During the observations the primary was placed behind the occulting disk.  The first 2012 data set used the astrometric grid spots \citep{sivaramakrishnan2006}, allowing the measurement of the astrometry of the companion even though the primary is occulted by the coronagraph. Two nights later on 2012 June 14 UT, we collected an additional data set, but this time the astrometric grid spot was not used and we were unable to measure the astrometry of this data set. Our final observation was on 2015 March 31 UT and it used the astrometric grid spots. All the data were reduced using the  Project 1640 data reduction pipeline described in \citet{zimmerman2011}.  Since then, a few upgrades have been made to the pipeline that reduce lenslet-lenslet cross talk.  The pipeline processing produces an image of the object in each of 32 wavebands, resulting in a data cube.

The astrometric grid spots produce four spots on each frame of the data cube.  The location of the spots is dependent on the wavelength of the frame.  To measure the astrometry  we fit a Gaussian to each of the four astrometric grid spots. Then we identified the location of the primary by fitting lines to the vertical and the horizontal spots.  The primary is located where these two lines intersect.   The final astrometry is the average of all the astrometry measured from 29 of the data slices from both of the image cubes with grid spots. We only use 29 image slices because several images in the atmospheric water bands did not produce meaningful results due to poor signal.   The plate scale and position angle offset of the detector were computed from observations of calibration binaries taken during each observing run. The astrometry is presented in Table \ref{astrometry}. 

\subsection{PHARO Observations}

We also collected near-IR images of the system on multiple dates using the PHARO near-infrared camera \citep{hayward2001}. Like P1640, PHARO is also mounted on the PALM-3000 AO system on the 5~m Hale telescope.  We collected data on 29 August 2013 UT, 14 May 2014 UT and  31 March 2015 UT. All observations used the 25\arcsec~field of view with  a pixel scale of 25 mas pixel$^{-1}$.  The observations used a variety of near-IR filters. After collection, each frame was debiased, sky subtracted, and flat fielded. We again used the $fitstars$ algorithm to measure the astrometry and photometry of the binary and assigned error bars using the technique described in \citet{roberts2005}

We also examined archival PHARO observations.  \citet{carson2005} observed $\mu$~Her in 2002 June with PHARO in coronagraphic mode in \textit{Ks} filter. It was not originally detected with the automated detection algorithm  due to nearby residual light from the central star's characteristic ``waffle pattern'', which the algorithm interpreted as a high noise level in that localized region of the image.  A visual inspection of the archival image  clearly shows the companion. The companion's relative astrometry was extracted by visual inspection.  Uncertainties were dominated by the ability to determine the center star's position behind the coronagraphic mask.  We conservatively assigned a 2 pixel (0\farcs08) uncertainty in the center star's position (see discussion in \citealt{carson2005}).       

\subsection{NIRSPEC/SCAM Observations}

\begin{figure}[thb]
    \centering
\includegraphics[width=4cm, height=4cm]{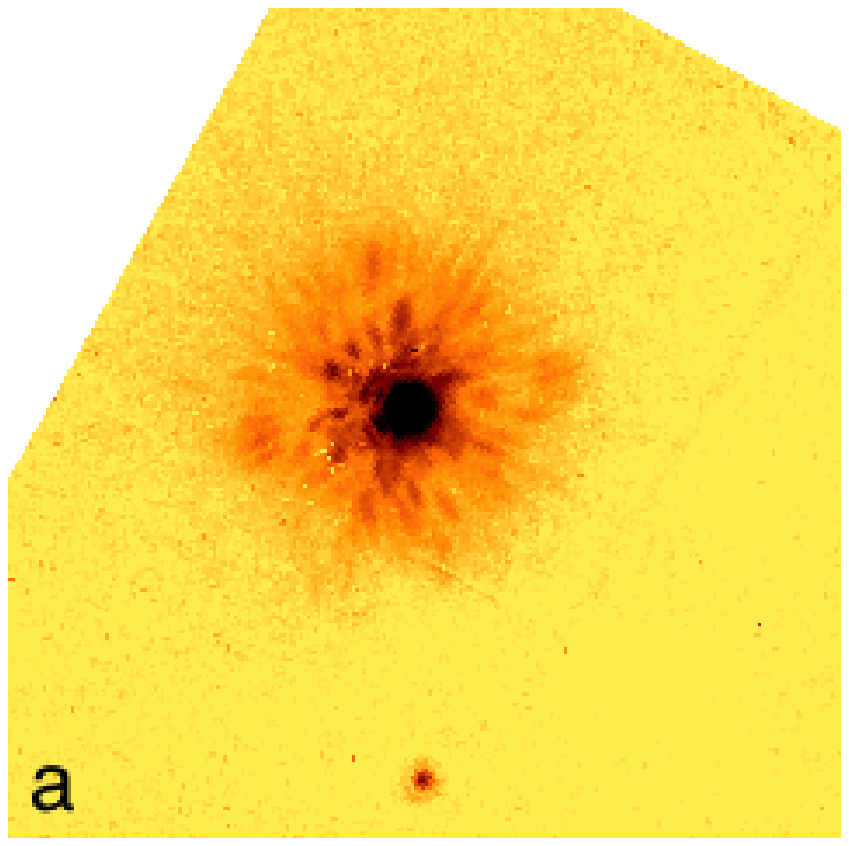} 
\includegraphics[width=4cm, height=4cm]{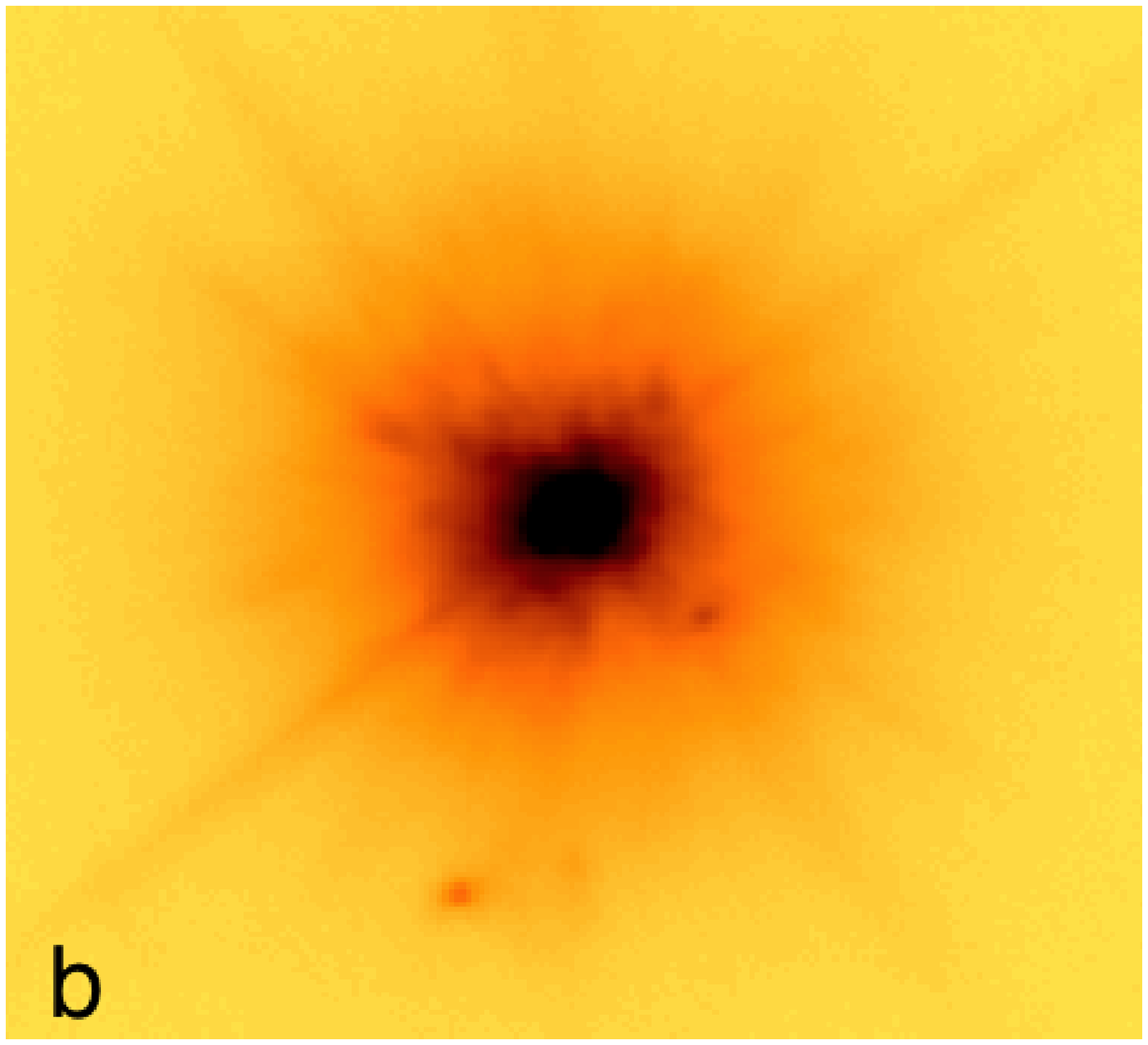} 
\includegraphics[width=4cm, height=4cm]{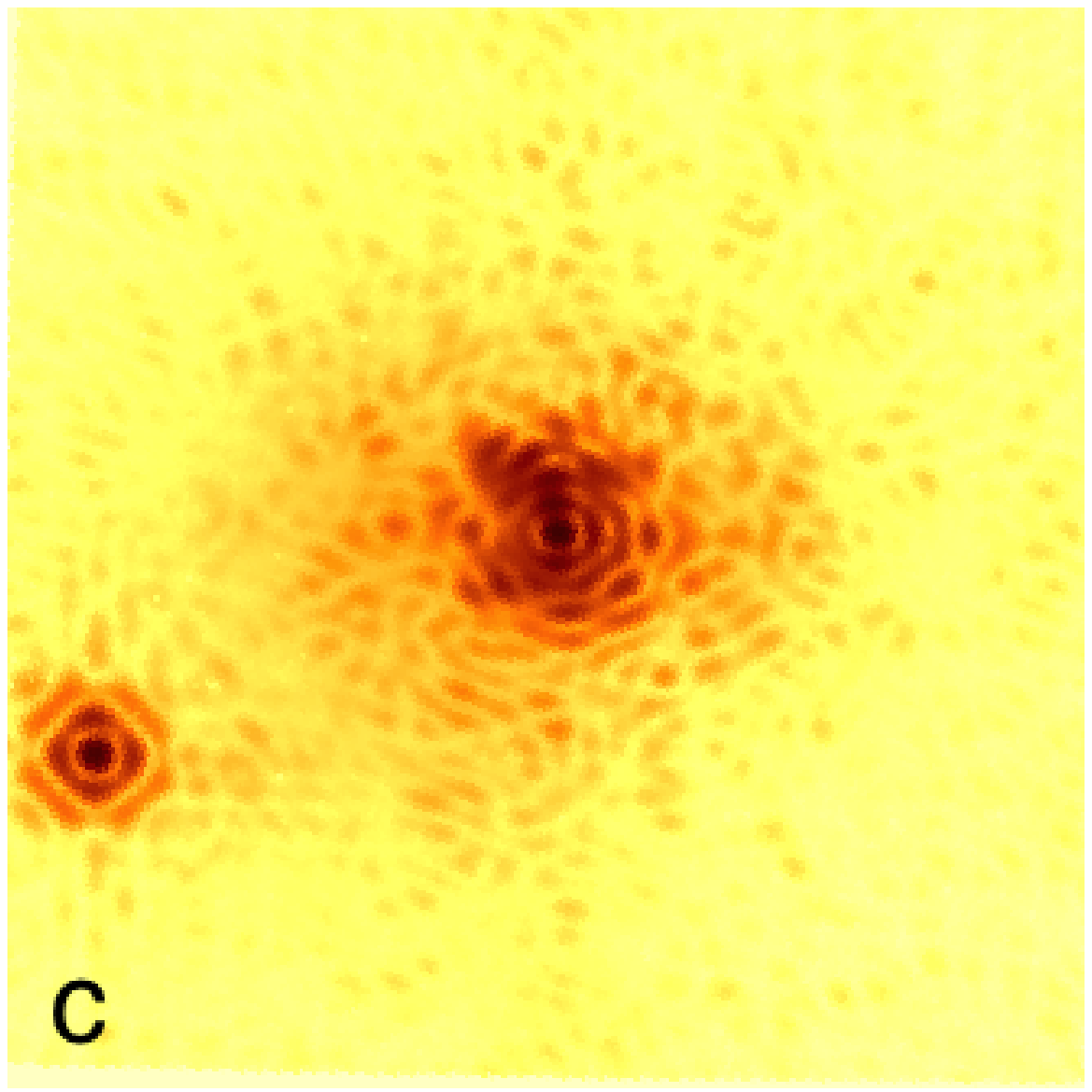} 
\includegraphics[width=4cm, height=4cm]{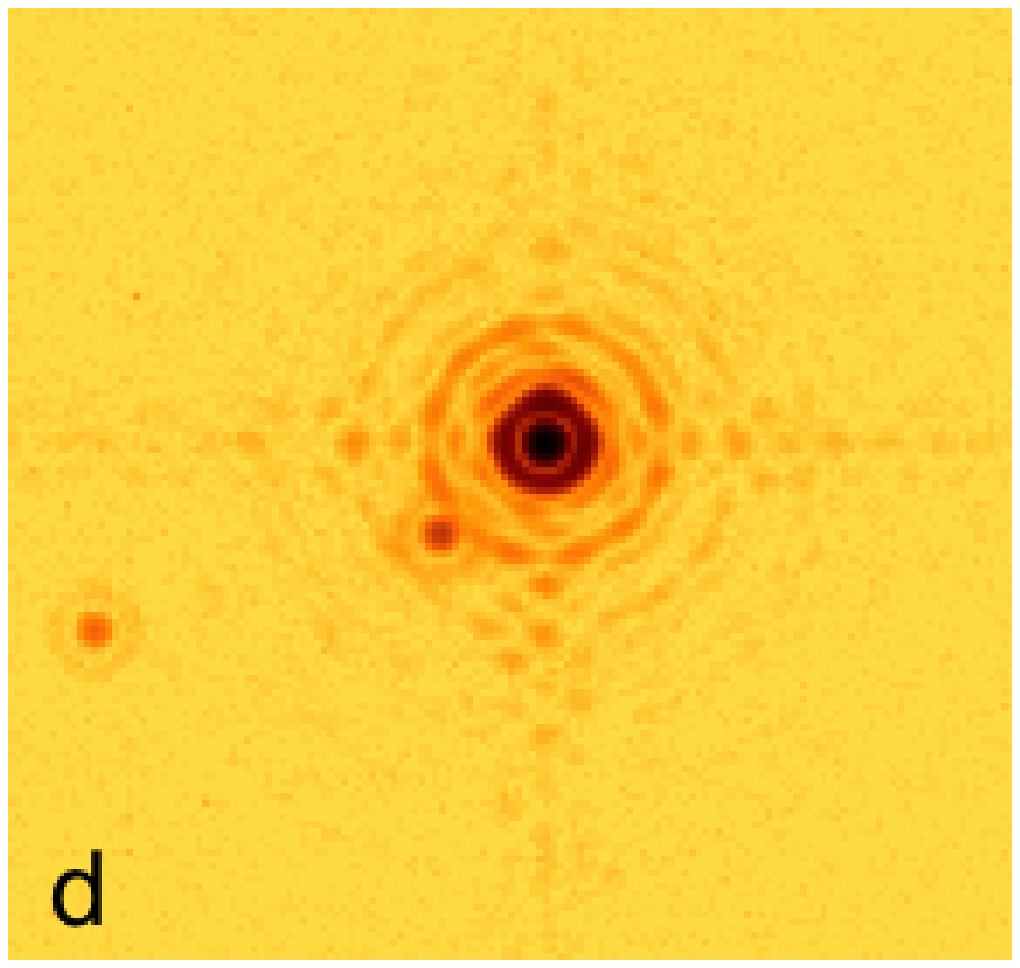} 
    \caption{Images of $\mu$~Her Aa,Ab  showcasing an image from each of the instruments used in this paper.  The instruments and dates of the images are: a) NIRSPEC/SCAM from 2001\footnote{Image was retrieved from the Keck Archive: https://koa.ipac.caltech.edu. PI of the original observation was M. Liu}, b) VisIm from 2005, c) P1640 from 2012, and d) PHARO from 2013.    All images have been rotated so that North is up and East is to the right.  Orbital motion of $\mu$~Her Ab is clearly seen as it rotates clockwise.  Each image has a horizontal size of approximately 3--4\arcsec.    The P1640 image is an occulted image and is a single slice of the image cube at a wavelength of 1.720\micron. In the PHARO image, the close object to the lower left of the primary is a ghost from a neutral density filter in the camera. }%
    \label{images}%
\end{figure}

We located images of $\mu$~Her Aa,Ab in the archives of the Keck II telescope.  The Keck II data were collected with NIRSPEC's Slit Camera (SCAM) \citep{mclean2000}  in AO mode  on 2001 May 15 UT with the NIRSPEC-7 filter with a central wavelength of 2.222~\micron~and a FWHM of 0.805 \micron.    We were unable to identify any calibration data for SCAM for the 2001 data. In the images the star was placed in different positions across the image.  To calibrate the images, we fit a plane to each image and subtracted it off. We then subtracted off a median of all the deplaned images. This resulted in 12 images. We then measured the astrometry of each image using \textit{fitstars}.  One of the images is shown in Figure \ref{images}a.  We set the error bars equal to the standard deviation of the measurements.   The resulting astrometry and photometry are given in Table \ref{astrometry}.

\begin{deluxetable*}{lccccl}
\tablewidth{0pt}
\tablecaption{Measured astrometry and photometry\label{astrometry}}
\tablehead{ \colhead{Epoch} & \colhead{$\theta$ ($^{\circ}$)} &\colhead{$\rho$ (\arcsec)} & \colhead{$\Delta$M} & \colhead{Filter}& \colhead{Instrument/Reference} }
\startdata
1998.5359   &   \phd\phn\phn167$\pm$3\phd\phn\phn      & \phn\phn1.42$\pm$0.04\phn       &7.26$\pm$0.15 & \textit{I} & \citet{turner2001}\\
1998.5360   &   \phd\phn\phn164$\pm$3\phd\phn\phn      & \phn\phn1.43$\pm$0.04\phn       &9.29$\pm$0.15 & \textit{R} & \citet{turner2001}\\
2000.38     &  \phn178.1$\pm$0.3\phn  &  \phn\phn1.38$\pm$0.007  &  5.0$\pm$0.1 & \textit{Br$\gamma$} & \citet{lloyd2002}\\
2000.77     &  \phn177.7$\pm$0.3\phn  &  \phn\phn1.35$\pm$0.007  &  5.3$\pm$0.1 & \textit{Br$\gamma$} & \citet{lloyd2002}\\
2001.3678   &   \phd\phn\phn182$\pm$0.3\phn   & \phn\phn1.40$\pm$0.008     & 5.59$\pm$0.13 &  N-7 & NIRSPEC/SCAM\\
2001.80\tablenotemark{a}     &    \phd\phn\phn184$\pm$8\phd\phn\phn     & \phn\phn\phn1.3$\pm$0.2\phn\phn         & \phn7.3$\pm$0.1\phn & \textit{H} & \citet{debes2002}\\
2001.80\tablenotemark{a}     &    \phd\phn\phn184$\pm$8\phd\phn\phn     & \phn\phn\phn1.3$\pm$0.2\phn\phn         & \phn6.8$\pm$0.1\phn & \textit{K} & \citet{debes2002}\\
2002.4622   &    \phd\phn\phn190$\pm$2\phd\phn\phn     & \phn\phn1.44$\pm$0.08\phn       & ...           &\textit{Ks} & PHARO\\
2003.5112   & \phn196.8$\pm$1.0\phn     & \phn\phn1.37$\pm$0.01\phn       & \phn7.1$\pm$0.2\phn   & \textit{I} & VisIm\\
2005.3126   & \phn207.5$\pm$1.0\phn     & \phn\phn1.39$\pm$0.01\phn       & \phn7.2$\pm$0.2\phn   & \textit{I} & VisIm\\
2006.28070  &  214.55$\pm$0.20    & 1.4907$\pm$0.005  & ...           & \textit{M} & \citet{kenworthy2007}\\
2006.28072  &  214.69$\pm$0.18    & 1.4847$\pm$0.005  & 4.99$\pm$0.06 & \textit{M'} &  \citet{kenworthy2007}\\
2012.4473   &  244.66$\pm$0.14    &1.718\phn$\pm$0.005  &   ...         &   \textit{yJH}  &  P1640 \\
2013.7426   &  \phn248.1$\pm$0.3\phn &\phn\phn1.77$\pm$0.01\phn      & 5.43$\pm$0.1\phn  & \textit{Br$\gamma$} & PHARO\\
2014.3669   &  \phn250.8$\pm$0.5\phn &\phn\phn1.77$\pm$0.01\phn      & 5.98$\pm$0.5\phn    & \textit{J} & PHARO\\
2014.3669   &  \phn250.6$\pm$0.5\phn &\phn\phn1.77$\pm$0.01\phn      & 5.67$\pm$0.5\phn    & \textit{H} & PHARO\\
2014.3669   &  \phn250.8$\pm$0.5\phn &\phn\phn1.79$\pm$0.01\phn      & 5.73$\pm$0.5\phn    & \textit{CH$_4$ Short} & PHARO\\
2014.3669   &  \phn251.1$\pm$0.5\phn &\phn\phn1.74$\pm$0.01\phn      & 5.33$\pm$0.1\phn    & \textit{Kcont} & PHARO\\
2014.3669   &  \phn251.1$\pm$0.5\phn &\phn\phn1.77$\pm$0.01\phn      & 5.43$\pm$0.1\phn    & \textit{Ks} & PHARO\\
2014.3669   &  \phn250.0$\pm$0.5\phn &\phn\phn1.76$\pm$0.01\phn      & 5.39$\pm$0.1\phn    & \textit{Br$\gamma$} & PHARO\\
2015.2432   &  \phn254.0$\pm$0.5\phn  &\phn\phn1.77$\pm$0.01\phn      & 5.44$\pm$0.1\phn    &\textit{Br$\gamma$} & PHARO\\
2015.2433   &  \phn254.0$\pm$0.5\phn  &\phn\phn1.77$\pm$0.01\phn      & 5.96$\pm$0.5\phn    &\textit{J} & PHARO\\
2015.2433   &  \phn253.7$\pm$0.5\phn  &\phn\phn1.76$\pm$0.01\phn      & 5.67$\pm$0.5\phn    &\textit{CH$_4$ Short} & PHARO\\
2015.2459   &  \phn254.8$\pm$0.1\phn  &\phn\phn1.78$\pm$0.01\phn      &   ...         &   \textit{yJH}  &  P1640 
\enddata
\tablenotetext{a}{The original paper does not list the observation date, by examining the original observing logs we determined that the observation date was in the range of 2001 October 15-21.}
\end{deluxetable*}

\section{ANALYSIS}\label{analysis}

\subsection{Spectral Type Determination}\label{spectral_type}

To create a spectrum of the stellar companion, we performed aperture photometry on each of the 32 images in the P1640 data cubes. This was done with the \textit{aper.pro} routine which is part of the IDL astrolib\footnote{http://idlastro.gsfc.nasa.gov} and is an adaptation of \textit{DAOphot} \citep{stetson1987}.  A photometry aperture and a sky annulus were centered on the companion.  The radii of the apertures were set equal to 3.5$\lambda/D$ rounded to the nearest pixel size, where $\lambda$ is the central wavelength of each image and  $D$ is the aperture of the telescope.   The annulus size was set to avoid the primary star and the occultation spot.    The spectrum is the measured power in each slice as function of wavelength. 

The extracted spectrum is a convolution of the object spectrum, the instrumental spectral response function (SRF) and the  atmospheric SRF \citep{roberts2012}. In some of our previous results, we have computed the combined SRF of the atmosphere and instrument  from unocculted images of the primary \citep{roberts2012}. $\mu$~Her A is so bright that it saturates our detector on even the shortest exposures. We computed the SRF from another star. On 2012 June 12 UT, there were no other suitable stars to extract an SRF,  so we used the SRF extracted from HD 129814 on 2012 June 14 UT. The airmass of the $\mu$~Her observations on 12 June was 1.02, while the airmass of HD~129814 was 1.04.  On 2012 June 14 UT,  $\mu$~Her was observed at an airmass of 1.01, and again we used the SRF of HD~129814. On 31 March 2015 UT, $\mu$~Her was observed at an air mass of 1.01.  We used the SRF computed from HD~74360 from the same night with an airmass of 1.02.  

After the spectrum of the companion was extracted from the P1640 data, we compared it against the spectra in the Infrared Telescope Facility (IRTF) Spectral Library \citep{cushing2005, rayner2009}. These include FGKM main sequence stars and  LT brown dwarfs. The template spectra were binned and smoothed in order to produce the equivalent spectra to having the star observed by P1640.   Then each template spectrum was compared against the measured spectrum using the sum of the squares of the residual (SSR) as a metric \citep{roberts2012}. The best fit reference spectrum was the one with the minimum value to the SSR.

\begin{figure*}[thb]
  \begin{center}
  \includegraphics[width=140mm]{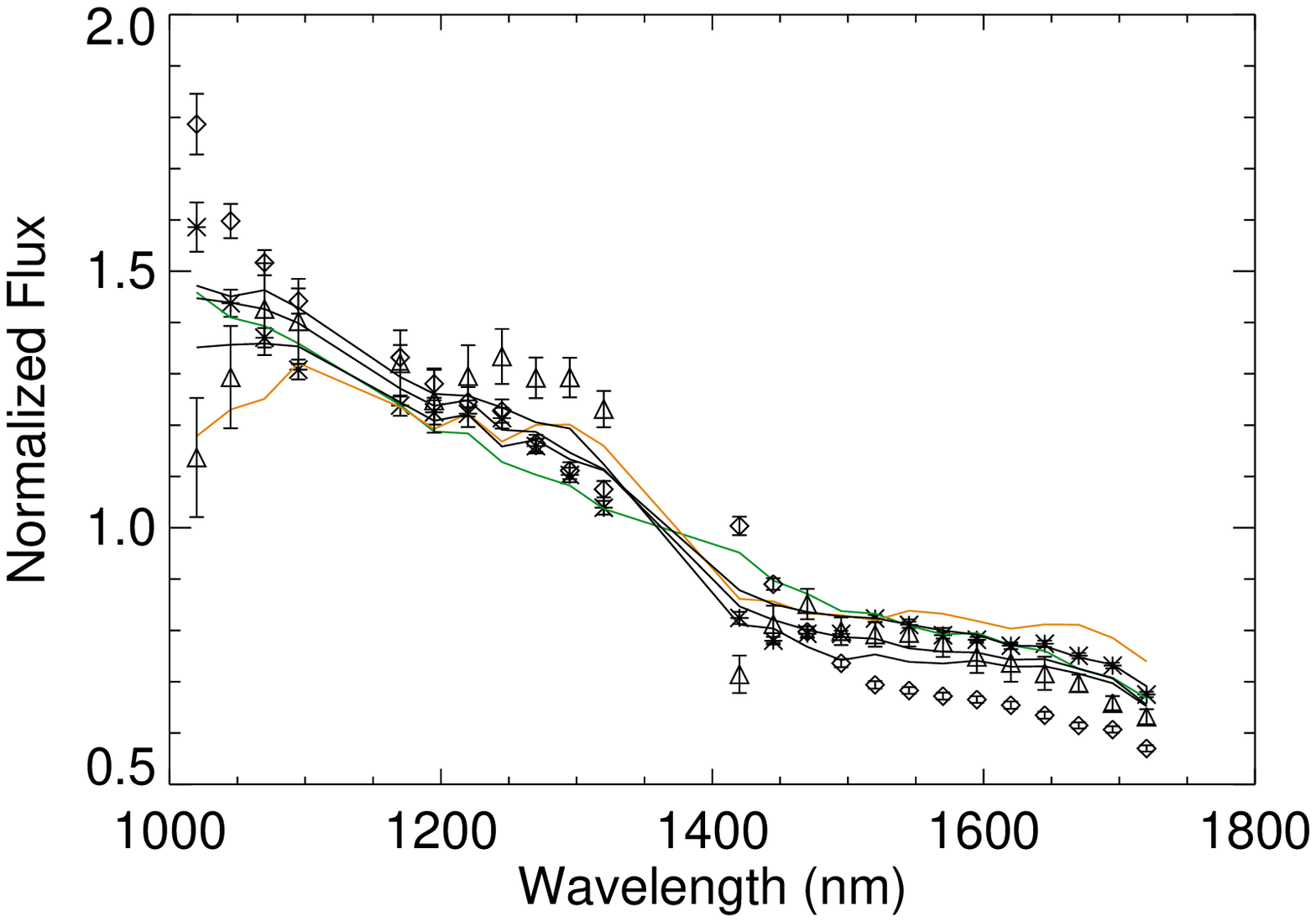}
   \end{center}
 \caption{The extracted spectra of $\mu$~Her Ab from P1640. The asterisks are the 2012 June 12 UT data,  the diamonds are the 2012 June 14 UT data and the triangles are the 2015 March 31 UT data.  The over plotted template spectra are Green=M2V, Black=M4V and Orange=M6V. There are three separate M4V spectra, to show the variance between spectra of the same spectral type. }
 \label{spectra}
\end{figure*}

\begin{figure}[thb]
    \centering
    \includegraphics[height=6cm]{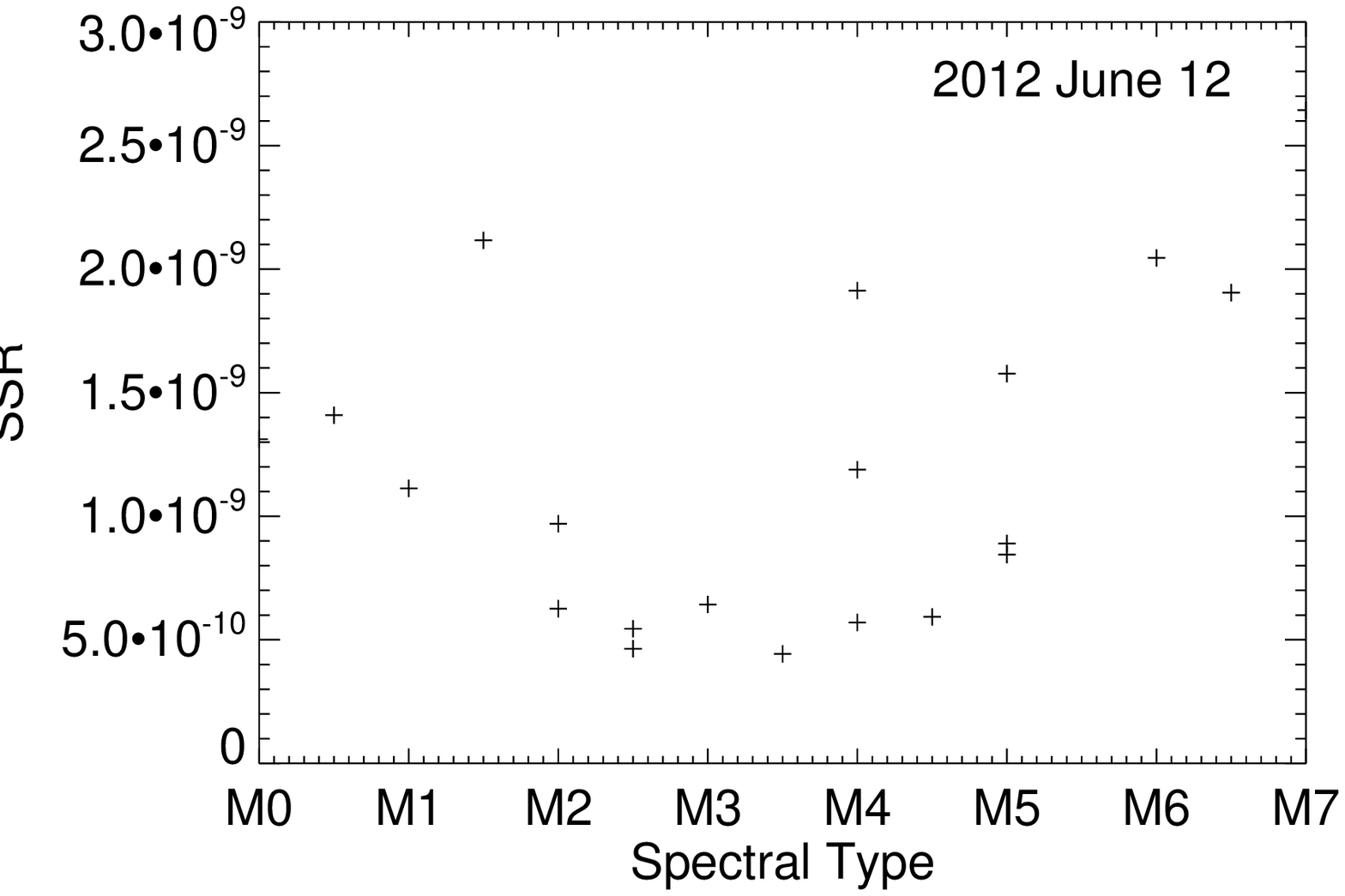} 
    \includegraphics[height=6cm]{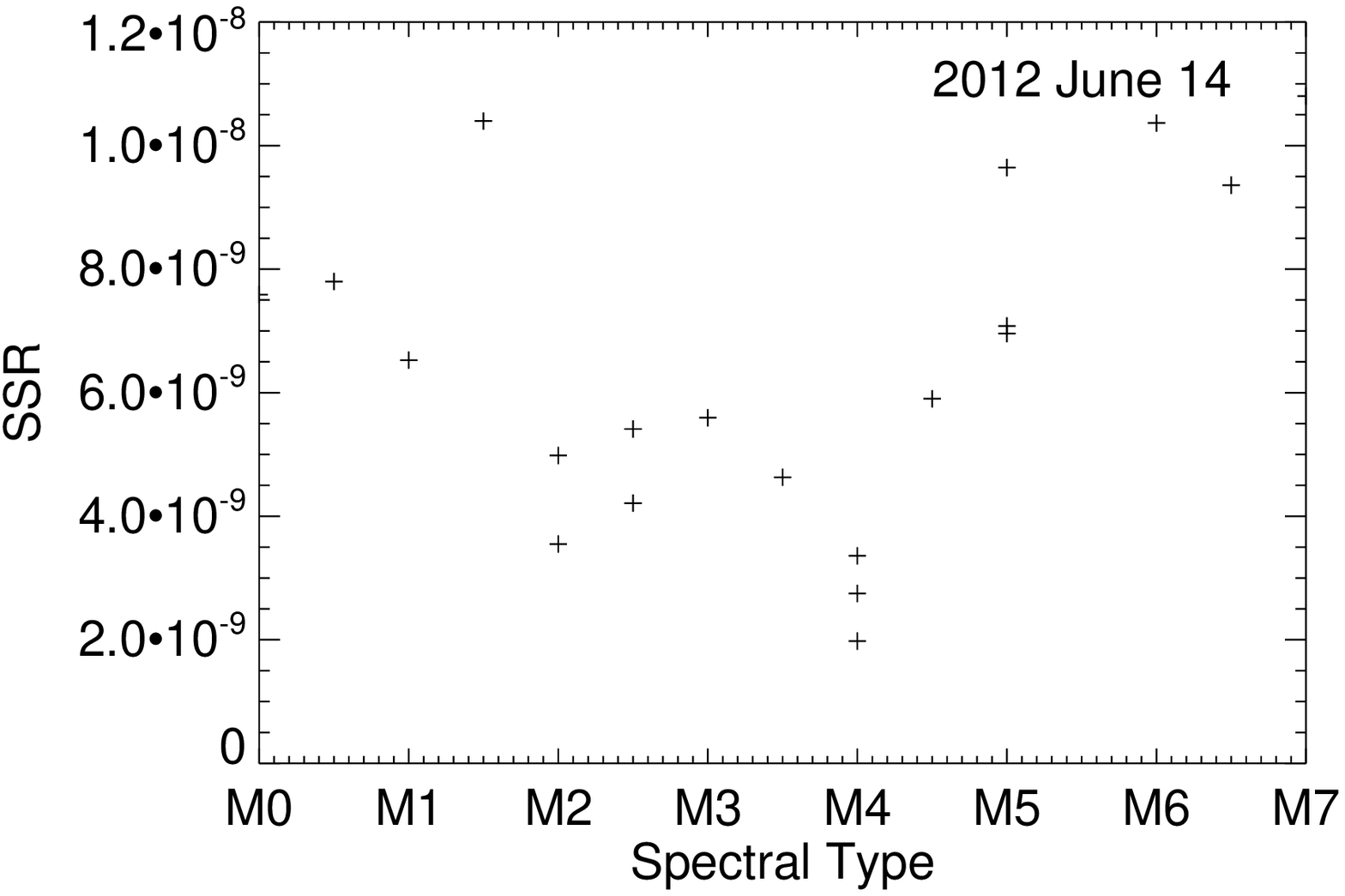} 
    \includegraphics[height=6cm]{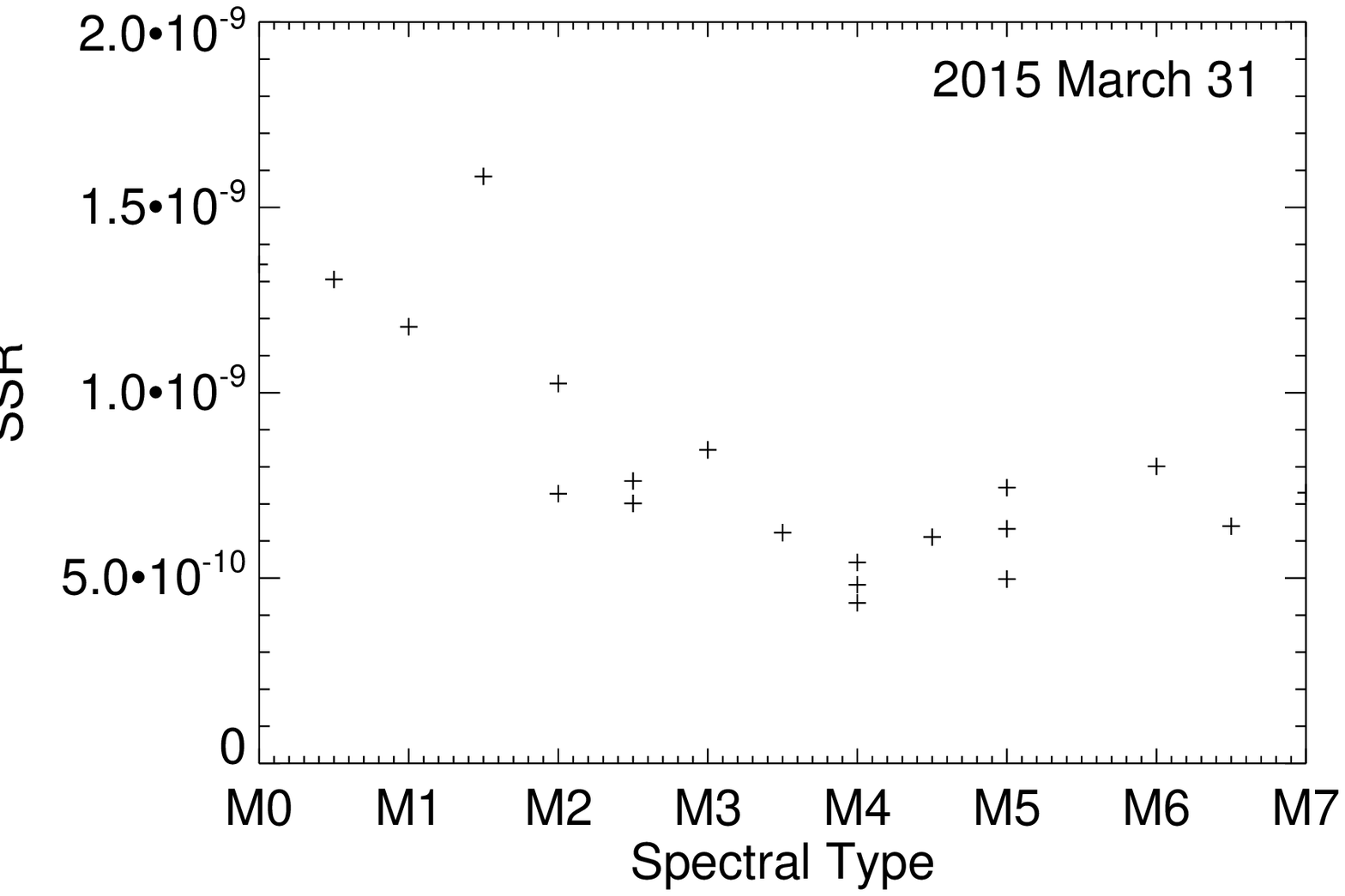}  
    \caption{The SSR metrics plotted as a function of spectral type for each of the three observations.   Each subfigure is labeled with the UT date of the observations.  }
    \label{metric}
\end{figure}

In Figure \ref{spectra} we plot the normalized spectra for  $\mu$~Her Ab for the three observations overplotted with the spectra for M2V, M4V, and M6V stars taken from the IRTF Spectral Library.  There are three M4V stars in the Spectral Library and we have plotted all three of them on this chart. This shows the variation between spectra of the same type due to age and metallicity. Figure \ref{metric} shows the sum of the squares of the residual (SSR) metric as a function of spectral type for each of the three observations. For spectral types having multiple entries in the IRTF Spectral Library, multiple values of the SSR metric are calculated and shown.  The 2012 June 12 SSR metric indicates the star has a spectral type of M2V-M5V.   The 2012 June 14 data have the highest value of the SSR, indicating the worst fit overall, but  the SSR metric strongly prefers the M4V result.  The 2015 March 31 result is a M4V-M5V.  From these three results, we conclude that $\mu$~Her Ab has a spectral type of M4$\pm$1V.

\subsection{Orbital Analysis}\label{orbit_analysis}

There have been a number of prior mass estimates of the primary star. \citet{fuhrmann1998}  derived a mass of 1.14 $M_\sun$ with uncertainties of 5\% by comparing the measured effective temperature against the luminosity on theoretical evolutionary tracks.   Using the same technique, \citet{takeda2005}  derived a mass of 1.13 $M_\sun$.  \citet{jofre2015} compared measured spectra to models to compute a mass of 1.09$\pm$0.01 $M_\sun$.  \citet{bonanno2008} detected solar-like oscillations in the primary with asteroseismology. \citet{yang2010} modeled these oscillations and  computed a mass of $\mu$~Her A of $1.00_{-0.02}^{+0.01}$ $M_\sun$ and an age of 6.433 $\pm$ 0.04 Gyr.  The authors note that models with mass 1--1.1 $M_\sun$ and age 6.2--6.7 Gyr can also reproduce the non-asteroseismic and asteroseismic constraints.   There have been two measurements of the primary's mass with long baseline interferometery.   \citet{boyajian2013} calculated a mass of 1.118 $M_\sun$.  \citet{baines2014} produced a estimate of 0.87$\pm$0.02 $M_\sun$.  With the exception of the \citet{baines2014} analysis, the results are all approximately 1.1 $M_\sun$ and we have used this value below.

In addition to the astrometry shown in Table \ref{astrometry}, we also have radial velocity  (RV) data  from the Lick Planet Search \citep{fischer2014}.  $\mu$~Her A was monitored from 1987 till 2013.   Over this time period, 332 RV data points were collected. The data have a median error of 2.26 m/s.  While not tracing out a complete orbit, the data clearly show acceleration.    We generated mean points for the astrometric data taken the same day.  We computed a combined orbital solution using the ORBITX\footnote{\url{http://www.ctio.noao.edu/$\sim$atokovin/orbit/index.html}} code \citep{tokovinin1992}. ORBITX uses the Levenberg-Marquardt method to solve for all the orbital elements.  

\begin{figure}[htb]
  \begin{center}
   \includegraphics[width=70mm]{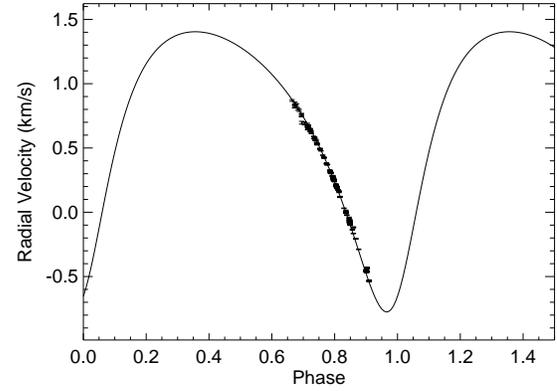}
  \end{center}
 \caption{The RV orbit of of $\mu$~Her Aa,Ab. The black line is the computed orbit, while the individual data points and their associated error bars are overplotted.  }
 \label{rv}
\end{figure}

\begin{figure}[htb]
  \begin{center}
  \includegraphics[width=70mm]{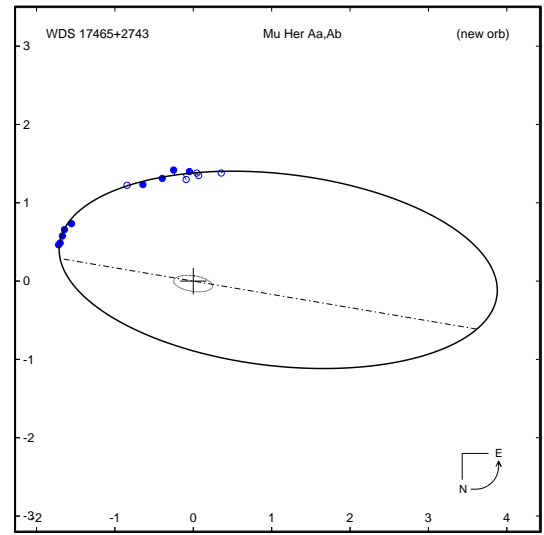}
   \end{center}
 \caption{Preliminary orbit of $\mu$~Her Aa,Ab.   The solid thin line orbit is the astrometric orbit from measuring the shift in the photo-center from \citet{heintz1994}.  The broken line through the origin is the line of nodes. Open circles are previously published astrometry values and the filled circles are the new values from this paper. The axes are labeled in units of arcseconds.}
 \label{visual_orbit}
\end{figure}

Figure \ref{visual_orbit} illustrates the new visual orbital solution, plotted together with all published data as well as the new data in Table \ref{astrometry}. Previously published observations are indicated by open circles, and measures from Table \ref{astrometry} are represented by filled circles.  ``$O-C$" lines connect each measure to its predicted position along the new orbit (shown as a thick solid line). A dot-dash line indicates the line of nodes, and a curved arrow in the lower right corner of each figure indicates the direction of orbital motion in addition to providing the figure orientation. The scale of the orbit is indicated. Finally, the previously published orbit of \citet{heintz1994} is shown as a dashed ellipse.  Figure \ref{rv} shows the computed spectroscopic orbit and the  archival RV data are overplotted.  The RV data include error bars, but they are hard to see due to the tiny errors on each data point. 

Table \ref{orbital_elements}  lists the solution for the nine orbital elements: $P$ (period, in years), $a$ (semi-major axis, in arcseconds), $i$ (inclination, in degrees), $\Omega$ (longitude of nodes, equinox 2000.0, in degrees), $T_0$ (epoch of periastron passage, in fractional Julian year), $e$ (eccentricity), $\omega$ (longitude of periastron, in degrees) K1 (semiamplitude of the primary,  in kms$^{-1}$) and V0 (the systemic velocity, in kms$^{-1}$).  For comparison purposes, the table also lists the orbital elements from the two previous astrometric orbits for the system from \citet{heintz1987} and \citet{heintz1994}, with the exception of semi-major axis: since the previous orbits were based on the motion of the photocenter, the derived semi-major axis from these previous studies is not expected to be the same.  There were no error bars on the elements from either of the Heintz orbits. \citet{heintz1987} estimated that three quarters of the orbit had been observed, which would be about 45 years.  The three orbits have consistent solutions for the inclination, longitude of node and eccentricity.  We have found a longer period than the previous orbits, but that is not unusual as astrometric orbits often have large errors. The orbital elements are most appropriately characterized as provisional. The solution should be good enough to provide  reasonable ephemerides over the next decade, but the elements will require correction over the course of a complete orbit to be  considered approximately correct. Despite the uncertainty, the quantity 3 $\log(a) - 2 \log(P)$, should not be grossly erroneous and provides useful information on the mass of the system.

Using the combined orbit and the measured parallax \citep{vanLeeuwen2007}, we are able to compute a mass sum of of $\mu$ Her Aa,Ab as 1.42 $M_\sun$ with a minimum mass of the secondary of 0.21$M_\sun$.   Using the previously derived mass of the primary from the literature of 1.1$M_\sun$ produces an estimate of the secondary mass of 0.32 $M_\sun$.  The secondary spectral type determined in \S \ref{spectral_type} corresponds to a mass of 0.14-0.36$M_\sun$ \citep{reid2005}, which is in general agreement with our estimate of the secondary mass from the combined orbit solution. 
 
\begin{deluxetable}{lccc}
\tabletypesize{\footnotesize}
\tablewidth{0pt}
\tablecaption{Orbital elements\label{orbital_elements}}
\tablehead{\colhead{Element}      &  \colhead{\citet{heintz1987}}  & \colhead{\citet{heintz1994}} & \colhead{This paper}}
P (yr)       & 63\phd   &  65  &  98.9$\pm$22.7\\
a (\arcsec)  & - & - &  2.9$\pm$0.3\\
i (\degr)    & 68\phd & 68 & 62.82$\pm$4.66\\
W (\degr)    & 80.0 & 81.8 & 80.4$\pm$1.7\\
T            & 1951.0& 1951.0 & 1921.1$\pm$23.8\\
e            & 0.34 & 0.32 & 0.44$\pm$0.06\\
w (\degr)    & 94.2 & 92 & 214\phd$\pm$16\phd\\
K1 (kms$^{-1}$) & - & - & 1.12$\pm$0.10\\
V0 (kms$^{-1}$) & - & - & 0.73$\pm$0.10
\enddata
\end{deluxetable}

\section{Discussion and Conclusions}\label{summary}

We have shown that $\mu$~Her Ab  is a stellar companion with a spectral class of M4$\pm$1V.  This was estimated from multiple epochs of low-resolution near-infrared spectra of the companion.  In addition we have computed a combined visual and spectroscopic orbit for the system, producing a mass estimate for the secondary which is in agreement with the expected mass of a M4V star.  

Much of the previous confusion about the stellar versus sub-stellar nature of $\mu$~Her Ab comes from combining photometric results from different observations.  While that is normally a fine practice, in this case it proved to be a problem.  The near-infrared photometry of \citet{debes2002} appears to be incorrect as it does not agree with the numerous PHARO observations in H and Ks nor the NIRSPEC/SCAM observation at 2.2 \micron.  Yet their estimate of the spectral type of the companion from H-K color agrees with ours from near-infrared spectroscopy.  It may be that there was an offset in the photometry from both filters.  This would also explain the anomalous $K-M$ color computed by \citet{kenworthy2007} from their mid-IR data combined with the \citet{debes2002} near-infrared photometry, which they reported as too red for a M4V star.  

We have computed a preliminary orbit for the $\mu$~Her Aa,Ab binary.  A continuing effort is needed to collect additional astrometric and radial velocity measurements of the system to improve the orbit determination. Additional RV data in the near future would be especially useful as the system has most likely just undergone an inflection point in the RV curve.  Additional data will produce a more accurate estimate of the masses of the two stars.  The mass of the primary can then be compared against the mass of the primary measured from asteroseimology.  Combining independent observations with asteroseismology is crucial to advance progress in the theoretical modeling of observed oscillation frequencies, and the validation of asteroseismic relations to derive fundamental stellar properties \citep{huber2014}.  


\acknowledgements

Our thanks to Nils Turner for assistance with the Mt.\ Wilson log books. We also thank Andrei Tokovinin for assistance with the latest version of his ORBITX code.  A portion of the research in this paper was carried out at the Jet Propulsion Laboratory, California Institute of Technology, under a contract with the National Aeronautics and Space Administration (NASA). This work was partially funded through the NASA ROSES Origins of Solar Systems Grant NMO710830/102190.  Project 1640 is funded by National Science Foundation grants AST-0520822, AST-0804417, and AST-0908484.  In addition, part of this work was performed under a contract with the California Institute of Technology funded by NASA through the Sagan Fellowship Program.  JA is supported by the Laboratory for Physical Sciences, College Park, MD, through the National Physical Science Consortium graduate fellowship program. J.C.\ was supported by the U.S. National Science Foundation under Award No.~1009203. R.N.\ was funded by the Swedish Research Council's International Postdoctoral Grant No.~637-2013-474. The members of the Project 1640 team are also grateful for support from the Cordelia Corporation, Hilary and Ethel Lipsitz, the Vincent Astor Fund, Judy Vale, Andrew Goodwin, and an anonymous donor. This paper is based on observations obtained at the Maui Space Surveillance System operated by the US Air Force Research Laboratory$'$s Directed Energy Directorate and at the Hale Telescope, Palomar Observatory. This research made use of the Keck Observatory Archive (KOA), which is operated by the W.~M.~Keck Observatory and the NASA Exoplanet Science Institute (NExScI), under contract with the National Aeronautics and Space Administration.  NExScI is sponsored by NASA's Origins Theme and Exoplanet Exploration Program, and operated by the California Institute of Technology in coordination with the Jet Propulsion Laboratory.   This research made use of the Washington Double Star Catalogue maintained at the U.S.~Naval Observatory, the SIMBAD database, operated by the CDS in Strasbourg, France and NASA's Astrophysics Data System.  
 
{\it Facilities:} \facility{AEOS (Visible Imager)}, \facility{Hale (Project 1640, PHARO)}, \facility{Keck:II (NIRSPEC)} 



\end{document}